\documentstyle[12pt]{article}

\parskip=\medskipamount
\begin{document}
\baselineskip=20pt

\begin{center}
{ \large \bf Disoriented Chiral Condensates and Anomalous Production of Pions
\footnote { Presented at the Workshop on Heavy Ion Collisions, Aug. 31.--Sept. 4, 1998, 
Bratislava, Slovakia }   }\\
\vspace{1.5cm}
{ \normalsize \bf M. Martinis, \footnote { e-mail address: martinis@thphys.irb.h
r}
 V. Mikuta-Martinis, \footnote {e-mail address: mikutama@thphys.irb.hr}
and J. \v Crnugelj }\\ 
\vspace{0.5cm}
Department of Physics, \\
Theory Division, \\
Rudjer Bo\v skovi\' c Institute, P.O.B. 1016, \\
41001 Zagreb,CROATIA \\
\end{center}
{ \large \bf Abstract}
\vspace{0.5cm}
\baselineskip=20pt

The leading-particle effect and the factorization 
property of the scattering amplitude in the impact parameter space 
are used to study semiclassical production of pions in the central 
region. The mechanism is related to the isospin-uniform solution of the nonlinear $\sigma $-model coupled to quark degrees of 
freedom. The multipion exchange potential between two quarks is 
derived. It is shown that the soft chiral pion bremsstralung also 
leads to anomalously large fluctuations in the ratio of neutral to 
charged pion.. 
We show that only direct production of pions in the form of an isoscalar 
coherent pulse  without isovector pairs can lead to large 
neutral-charged fluctuations.

\vspace{1cm}

PACS numbers: 25.75. + r, 12.38.Mh, 13.85.Tp, 24.60.Ky
\newpage
\baselineskip=24pt
\section{Introduction}

The old puzzle in cosmic-ray observations is the existence of  few                         
events characterized by an anomalously large  
number of charged pions in comparison with the number of neutral pions,
the Centauros [1], indicating that there should exist a  strong long-range correlation
between  two types of the pions. The negative results of the accelerator
searches for Centauros at CERN [2,3,4,5] and at Fermilab [6] 
suggest that the mechanism for their production is not yet well understood and that the production threshold must be larger then 1.8 TeV.

Such long-range correlations are possible if pions are produced 
semiclassicaly  and constrained by global conservation of isospin [7-12].

Although the actual dynamical mechanism of the
production of a classical pion field in the course
of a high-energy collision is not known, there exist 
numerous interesting recent theoretical attemps to explain Centauros either as 
different types of isospin fluctuations due to the  formation of a  
disoriented chiral condensate (DCC) [13-17], or as multiparticle Bose-Einstein 
correlations (BEC) [18], or as the formation of a strange quark matter (SQM) [19].
Among the most interesting speculations is the idea of DCC that localized                            
regions of misaligned chiral vacuum might occur during the
ultrahigh-energy hadronic and heavy-ion collisions.
These regions, if produced, would behave as a pion laser, relaxing to the 
ground state by coherent pion emission . It is  generally accepted that the 
fluctuation of the ratio of neutral to charged pions of the Centauro type could be 
a sign of the DCC formation provided that a single large domain is formed, containing 
a large number of low $p_{T}$ pions. Since the pions formed in the DCC are essentially 
classical they form a quantum superposition of coherent states with different orientation in
isospin space. If all the pions in the domain are pointing in the same isospin direction 
and the condensate state is a pure isoscalar then
the probability distribution of the ratio of neutral to total pions in the domain will be given by [10,13,14,16]
\begin{equation} 
P_{DCC}(f) = \frac{1}{2\sqrt{f}}
\end{equation}
where $f = n_{0}/n$ is the ratio of the number of $\pi _{0}$'s in the DCC divided by the 
total number of pions. The distribution $ P_{DCC}(f)$ is clearly very distinct from the binomial
distribution  which assumes equal probability for production of $\pi _{+}$, $\pi _{-}$ and  $\pi _{0}$  
pion  and is of the form
\begin{equation} 
P_{B}(n_{0},n) = {n \choose n_{0}}\left(\frac{1}{3}\right)^{n_{0}} \left(\frac{2}{3}\right)^{n -n_{0}}
\end{equation} 
which in the limit as $n\rightarrow \infty , n_{0}\rightarrow  \infty  $~ 
with  $ f$  fixed,  approaches a delta function at $f = 1/3$. 

The possibility of  observing the DCC type fluctuations  
critically depends on the size  and the energy content of the DCC domain. If this domain is of the pion size, 
the effect of DCC will be too small to be observed experimentally. Therefore the signals such as the isospin 
fluctuations, the enhancement of the number of low $p_{T}$ pions, and the suppresion of HBT 
correlations are all characteristics of a single large coherent emission domain  [20]. 
The space-time scenario of the formation and decay of the DCC is usually studied within 
one of the simplified versions of the chiral effective Lagrangians, either the linear or nonlinear 
sigma model [17,21]. However, the use of sigma  models,
be they linear or nonlinear, is only a rough approximation to the true situation, because the 
couplings of pions and sigma to the constituent quarks may be large and their effect should not be ignored. 
  
In the early models [7,8] the coherent
production of pions was taken for granted and considered as a dominant mechanism.
These models  also predict strong negative correlations between the number
of neutral and charged pions. In fact, it was observed long ago [9,10] that 
an uncorrelated jet model involving only pions with the global conservation of isospin 
gives the same pattern of neutral/charged fluctuations
as observed in Centauro events. This strong negative
neutral-charged correlation is a
general property of the independent pion emission in
which the cluster formation ( or the short-range correlation between pions )
is not taken into account [11,12,22,23]. What makes these older results interesting is the 
possibility that such a coherent isosinglet pulse of  pions could be produced in the high-energy 
collisions as a result of  de-excitation of a highly excited region of space, in which the 
chiral orientation of the order parameter -- the quark condensate -- is different from that of the
normal vacuum configuration [17,24].

In this paper, following the approach of our earlier papers [23],
 we consider in Section 2. the leading-particle effect as a  source 
of a classical pion field in the impact parameter space. Since the condition 
of approximately vanishing isospin of the coherent  pulse of pions is crucial for the observation 
of the DCC phenomena,  pions should  be produced in the nearly baryon-free 
central rapidity region .
 
In Section 3. we show how the factorization property of the
scattering amplitude in the impact parameter space may be related to the isospin-uniform solutions of 
the equation of motion of the  nonlinear sigma model in the presence of  a classical source 
containing constituent quarks [25]. We also study in Section 3. the quantum case of the nonlinear $ {\sigma} $  model and derive  the form of the multipion
exchange potential between two  quarks.

 Our results are summarized in the Section  4. Our general conclusion is that within the nonlinear $\sigma $ model the large isospin
fluctuations are consequence of singly produced  pions  which are constrained by global conservation of isospin.

\section{Pion production from a classical  source}  

\subsection{The eikonal $S$ matrix with isospin}

At high energies most of the pions are
produced in the central region. To isolate the central production,
we adopt the high-energy longitudinally dominated kinematics, with
leading particles retaining a large fraction of their incident
momenta. We assume that the collision energy is large enough
so that the central region is free of baryons.
The energy available for the hadron production is
\begin{equation}
E_{had} = \frac{1}{2} \sqrt{s} - E_{leading}
\end{equation}
which at fixed total c.m. energy $ \sqrt{s} $ varies from event to event.
Using the following set of $ ( 3 n + 2 ) $ independent variables $s, \; \{
\vec{q}_{iT},y_{i} \} \equiv q_{i}, \; \; i = 1,2, \ldots n,$
the n-pion contribution to the $s$-channel unitarity can be written as an
integral over the relative impact parameter  $b$ of the two
incident leading particles:
\begin{equation}
F_{n}(s) = \frac{1}{4s} \int d^{2}b \prod_{i=1}^{n}dq_{i}
\mid T_{n}(s, \vec{b};1 \ldots n) \mid^{2},
\end{equation}
where $ dq = d^{2}q_{T}dy/2(2 \pi)^{3}.$ The normalization is
such that
\begin{eqnarray}
F_{n}(s) & = & s \sigma_{n}(s), \nonumber \\
\sigma_{inel}(s) & = & \sum_{n=1}^{ \infty} \sigma_{n}(s).
\end{eqnarray}

The basic assumption of the independent pion-emission model,
in b-space is the factorization property of the
scattering amplitude $ T_{n}( s, \vec{b}; 1 \ldots n )$ :
\begin{equation}
T_{n}( s, \vec{b}; 1 \ldots n) = 2sf(s, \vec{b}) 
\prod_{\sigma=+,0,-}\prod_{i_{\sigma}=1}^{n_{\sigma}} 
\frac{ \textstyle i^{n_{\sigma}-1}}{\textstyle \sqrt {n_{\sigma}!}}
J_{\sigma}( s, \vec{b}; q_{i_{\sigma}}),
\end{equation}
where, owing to unitarity,
\begin{equation}
\mid f(s, \vec{b}) \mid^{2} = e^{ \textstyle - \overline{n}(s, \vec{b})}
\end{equation}
and
\begin{equation}
\overline{n}(s, \vec{b}) = \sum_{\sigma}\int dq \mid J_{\sigma}(s, \vec{b};q) \mid^{2}
\end{equation}
denotes the average number of emitted pions at a given impact
parameter $b$.
The functions $ \mid J_{\sigma}(s, \vec{b}; q) \mid^{2}$, after the integration over b,
control the shape of the single-particle inclusive distribution.
A suitable choice of these functions also
guarantee that the energy and the momentum
are conserved on the average during the collision.
The factorization properties of $T_{n}$ in the form (6) is a consequence of
the pion field satisfying the following  equation of motion
\begin{equation}
( \Box + m_{\pi}^{2}) \vec{ \pi}( s,\vec{b}; x) = \vec{j}(s,
\vec{b};x),
\end{equation}
where $ \vec{j}$ is a classical source related to $ \vec{J}(s,
\vec{b};q )$ via the Fourier transform
\begin{equation}
\vec{J}(s, \vec{b};q) = \int d^{4}x e^{ iqx}
\vec{j}(s, \vec{b};x).
\end{equation}

The standard solution of Eq.(9) is usually given in terms of in- and
out-fields that are connected by the unitary
$S$- matrix $ \hat{S}(s, \vec{b};\vec{J})$ :
\begin{equation}
\vec{ \pi}_{out} = \hat{S}^{ \dagger} \vec{ \pi}_{in}
\hat{S} = \vec{ \pi}_{in} + \vec{ \pi}_{cl},
\end{equation}
where
\begin{equation}
\vec{ \pi}_{cl}(x) = \int d^{4}x' \Delta(x-x'; \mu) \vec{j}(s,
\vec{b};x').
\end{equation}
and $\Delta =\Delta_{ret} - \Delta_{adv}$.
The $S$ matrix following from such a classical source
is still an operator in the space of pions.
The coherent production of pions
is described by the following $S$ matrix:
\begin{equation}
\hat{S}(s, \vec{b};\vec{J}) = \int d^{2} \vec{e} \mid \vec{e}
\rangle D(s, \vec{b};\vec{J}) \langle \vec{e} \mid,
\end{equation}
where $ \mid \vec{e} \, \rangle $ represents the isospin-state
vector of the leading-particle system.
The quantity $D(s, \vec{b};\vec{J})$ is the unitary coherent-state
displacement operator defined as
\begin{equation}
D(s, \vec{b};\vec{J}) = exp[ \int dq
\vec{J}(s, \vec{b};q) \vec{a}^{ \dagger}(q) -h.c.],
\end{equation}
where $ \vec{a}^{ \dagger}(q)$ denotes the creation operator of
the pion field.

\baselineskip=24pt
\subsection{Distribution of pions in isospace}
If the isospin  of the incoming
leading particles is $II_{3}$ , then the initial-state
vector of the pion field is $ \hat{S}(s, \vec{b};\vec{J}) \mid II_{3} \rangle,$
where $ \mid II_{3} \rangle$ denotes a vacuum state with no pions.
The $n$-pion production amplitude is
\begin{equation}
iT_{n}(s, \vec{b};q_{1} \ldots q_{n}) = 2s \langle I'I'_{3};q_{1} \ldots
q_{n} \mid \hat{S}(s, \vec{b};\vec{J}) \mid II_{3} \rangle,
\end{equation}
where $ I'I'_{3}$ denotes isostate of the outgoing leading particles.

 If each possible isospin $(I', I_{3}')$ of the outgoing leading particle
system is produced with
equal probability, then we can sum over all $(I'I'_{3})$ using the group
theory alone to obtain the following probability distribution of producing $n_{+} \pi^{+}, \,
n_{ \_} \pi^{ \_},$ and $n_{0} \pi^{0}$:
\begin{eqnarray}
P_{I I_{3}}(n_{+}n_{ \_}n_{0}) N_{I I_{3}}& \! \! = \! \! & \sum_{I'I'_{3}}
\int d^{2}bdq_{1}dq_{2}
\ldots dq_{n} \mid \langle I'I'_{3}n_{+}n_{ \_}n_{0} \mid \hat{S}(s,
\vec{b};\vec{J}) \mid II_{3} \rangle \mid^{2}, \\
 n & \! \! = \! \! & n_{+} + n_{ \_} + n_{0} 
\end{eqnarray}
where $ N_{I I_{3}}$ is the corresponding normalization factor determined by \\
$ \sum_{\{n\}} P_{I I_{3}}(n_{+}, n_{-}, n_{0}) = 1 $.

This is our basic relation for calculating various pion-multiplicity
distributions, pion multiplicities, and pion
correlations between definite charge combinations.
In general, the probability $P_{I I_{3}}(n_{+}n_{ \_}n_{0})$
would depend on $(I'I_{3}')$ dynamically. The final-leading-particles
 tend to favor the $(I',I_{3}') \approx
(I,I_{3})$ case. However, if the leading particles are
colliding nuclei, an almost equal
probability for various $(I',I_{3}')$ seems reasonable aproximation
owing to the large number of possible leading isobars in the final state.
In the model leading to Eq.(16) the dynamical information which restricts the
possible values $(I'I'_{3})$ was not taken into account. Nevertheless, even if we assume that $(I',I'_{3}) \approx (I,I_{3})$ the pion's cloud  could
still have the isospin $ I_{ \pi} = 0,2, \ldots , 2I.$

Recent studies of heavy-ion collisions at the partonic level [26]
argue that the central region is mainly dominated by gluon jets. The
valence quarks of the incoming particles which
escape from the interaction region form the outgoing leading particle
system. Since gluon's isospin is zero, it is very likely that total
isospin of the produced pions in the central region is also zero. This
picture is certainly true if the central region is free from valence
quarks, the situation expected to appear at the extremelly high collision
energies.

If the conservation of
isospin is a global property of the colliding system, restricted only by
the relation
\begin{equation}
\vec{I} = \vec{I'} + \vec{I_{ \pi}},
\end{equation}
where $ \vec{I_{ \pi}}$ denotes the isospin of the emitted pion
cloud, then $ \vec{J}(s, \vec{b};q)$ should be of the form
\begin{equation}
\vec{J}(s, \vec{b};q) = J(s, \vec{b};q) \vec{e},
\end{equation}
where $ \vec{e}$ is a fixed unit vector in isospace independent of q.
The global conservation of isospin thus introduces the long-range correlation
between the emitted pions.

We shall analyze the isospin structure of the model
in the so called grey-disk approximation in which
\begin{equation}
\overline{n}(s, \vec{b}) = \overline{n}(s) \Theta(R(s)-b),
\end{equation}
where $ \overline{n}(s)$ denotes the mean total number of pions produced,
 $R(s)$ is related to the total inelastic
cross section, and $ \Theta$ is the step function.
For $ I_{3} = I,$ it is a straightforward algebra to calculate
the probability of creating $n$ pions of
which $n_{0}$ are neutral pions:
\begin{eqnarray}
P_{I}( n_{0} \mid n ) & = & \sum_{{n_{+}}+n \_
= n-n_{0}}P_{II}(n_{+}n \_ n_{0}) \\
& = & \left( \begin{array}{c}
n \\ n_{0} \end{array} \right) \frac{B(n_{0}+ \frac{1}{2},
n-n_{0}+I+1)}{B( \frac{1}{2},I+1)}.
\end{eqnarray}
Here $B(x,y) = \frac{ \textstyle \Gamma(x) \Gamma(y)}{
\textstyle \Gamma(x+y)}$
is the Euler beta function. Note that $P_{I}(
n_{0} \mid n)$ differs considerably from the binomial
distribution given here by $3^{-n}2^{n-n_{0}}( \stackrel{
\textstyle n}{n_{0}}).$

In Fig. 1 we show the  behavior of $P_{I}$
for $n=50$ and isospin $I=0,1$
of the initial-leading-particle system
in comparison  with the corresponding binomial distribution.
It is easy to see
that in the limit $ n \rightarrow \infty$, with $f=\frac{\textstyle n_{0}}{\textstyle n}$ fixed,
the probability distribution $nP_{I}(n_{0} \mid n)$
scales to the limiting behavior:
\begin{equation}
nP_{I}(n_{0} \mid n) \rightarrow P_{I}(f) =
\frac{(1 - f)^{I}}{B( \frac{1}{2},I+1) \sqrt{f}}.
\end{equation}
This limiting probability distribution is different from
the usual Gaussian random distribution for
which one expects $P_{I}(f)$ to be peaked at
$f = \frac{1}{3}$ as $n \rightarrow \infty.$
However, if pions are produced through the coherent
production of clusters which subsequentlly decay into two or more pions
the fraction of neutral pions in an event changes  substantially.

\subsection{Production of isovector clusters}

Let us assume that pions are produced both singly and through isovector
clusters of the $ \rho$ type[23]. In this case, the most appropriate tool for 
 studying pion correlations 
is the generating function $G_{II_{3}}(z,n_{ \_})$:
\begin{equation}
G_{II_{3}}(z,n_{ \_}) = \sum_{n_{0},n_{+}}P_{II_{3}}(n_{+},n_{ \_},
n_{0})z^{n_{0}},
\end{equation}
from which we can calculate, for example
\begin{eqnarray}
\langle n_{0} \rangle_{n_{ \_}} & = & \frac{d}{dz} \ln G_{II_{3}}
(1,n_{ \_}), \\ 
f_{2,n_{ \_}}^{0} & = & \frac{d^{2}}{dz^{2}} \ln G_{II_{3}}(1,n_{ \_}),
\end{eqnarray}
and
\begin{equation}
P_{II_{3}}(n_{0}) = \frac{1}{n_{0}!}\frac{ d^{n_{0}}}{dz^{n_{0}}} \sum_{n_{ \_}}
G_{II_{3}}(0, n_{ \_}). \\
\end{equation}
The form of this generating function in the case of $I = I_{3} = 0$ is
particularly simple:
\begin{equation}
G_{0 0}(z,n_{ \_})  = \int_{0}^{1} dx \frac{ \textstyle
\left[ A(z,x) \right]^{n_{0}}}{ \textstyle n_{0}!}e^{ \textstyle - B(z,x)},
\end{equation}
where
\begin{equation}
2A(z,x) = (1-x^{2}) \overline{n}_{ \pi}+z(1-x^{2}) \overline{n}_{ \rho}+
2x^{2} \overline{n}_{ \rho}
\end{equation}
and
\begin{equation}
2B(z,x) = \overline{n}_{ \pi}(1+x^{2} - 2zx^{2}) +
 \overline{n}_{ \rho} (2 - z(1-x^{2})).
\end{equation}
Here $ \overline{n}_{ \pi}$ denotes the average number of singly
produced pions, and $ \overline{n}_{ \rho}$ denotes the average number
of $ \rho$-type clusters which decay into two short-range correlated
pions.  Note that $A(1,x)=B(1,x)$.

The total number of emitted pions is
\begin{equation}
\overline n = \overline{n}_{ \pi} + 2 \overline{n}_{ \rho}.
\end{equation}

In Fig. 2. we show the behavior of $P (n_{0})\equiv P_{00}(n_{0})$ for
$ \overline n = 50$ and different combinations of $( \overline{n}_{ \pi}, 
\overline{n}_{ \rho}) $.  
We see that Centauro-type behavior is obtained only for $ \overline{n}_{ \pi} \neq 0$ and
$ \overline{n}_{ \rho}=0.$ Recent estimate of the ratio of
$ \rho$-- mesons to pions, at accelerator energies, is
$ \overline{n}_{ \rho} = 0.10  \overline{n}_{ \pi}$.
The behaviour of $\langle n_{0} \rangle_{n_{ \_}}$ and $ f_{2,n_{ \_}}^{0} $can be found in [23].

\subsection{Random source}

In this section we want to study the multiplicity distribution of pions emitted from a classical random source. In previous sections we have averaged our results only over a chiral orientations of the classical source but not the overall shape of the source.We assume that incident leading particles, which can neither be created nor destroyed, are acting as a classical random source for pions.

The unitary $ S $ matrix following from such a classical random source is $ \hat{S}(s, \vec{b};\vec{J})$.The initial-state vector
for the pion field is $\hat{S}(s, \vec{b};\vec{J}) \mid I I_{3} \rangle $.In practice, we rarely have any information about this initial state.It means that physical quantities should be averaged over the choice of source functions present in the initial state. In quantum statistics, the ensemble average is usually performed using the density operator, which, in our case, is of the form [27] 
\begin{equation}
\rho_{I I_{3}} (s, \vec{b}) = {[{\hat{S}(s, \vec{b};\vec{J})\mid I I_{3} \rangle \langle I I_{3} \mid {\hat{S}}^{\dagger}(s, \vec{b};\vec{J})}]}_{av}
\end{equation}
and normalized to unity: $ Tr[\rho_{I I_{3}}] = 1 $
In terms of the pion-number operator 
\begin{eqnarray}
N & = & N_{+} + N_{-} + N_{0} \nonumber\\
& = & \sum_{\sigma} \int dq a_{\sigma}^{\dagger}(q) a_{\sigma}(q)
\end{eqnarray}
the pion-generating function in the impact-parameter space becomes 
\begin{eqnarray}
G_{I I_{3}}(z_{0},z_{+},z_{-}) & = & Tr[ \rho_{I I_{3}}(s, \vec{b}) : e^{\sum_{\sigma}(z_{\sigma} - 1) N_{\sigma}} :] \\
& = & \langle exp [ \sum_{\sigma}(z_{\sigma} - 1) \overline{n}_{\sigma}(s,\vec{b})]\rangle _{I I_{3}} 
\end{eqnarray}
where 
\begin{equation}
\overline{n}_{\sigma}(s, \vec{b}) = \int dq \mid J_{\sigma}(s, \vec{b};q) \mid^{2}.
\end{equation}
The global conservation of isospin and the identity of pions require 
\begin{equation}
J_{\sigma}(s, \vec{b};q) = J(s, \vec{b};q) \vec{e_{\sigma}}
\end{equation}
where 
\begin{eqnarray}
e_{0} & = & e_{3} \\ 
e_{\pm} &  = & \frac{1}{\sqrt{2}}(e_{1}\mp i e_{2}).
\end{eqnarray}
Since 
\begin{equation}
\overline{n}_{\sigma}(s, \vec{b}) = {\mid e_{\sigma}\mid}^{2} \overline{n}(s, \vec{b})\\
\end{equation}
we can find the generating function in $ b $-space for producing  $ n_{ch} $ charged and $n_{0}$ neutral pions which is of the form
\begin{equation}
G_{II_{3}}(z_{0},z_{ch})  =  (I + \frac{1}{2})
 \frac{(I-I_{3})!}{(I+I_{3})!}
\int_{-1}^{1}dx \mid P_{I}^{I_{3}}(x) \mid^{2} \langle e^{\overline{n}(s, \vec{b}) z(x)}\rangle\\
\end{equation}
where 
\begin{equation}
z(x) = x^{2} z_{0} + (1- x^{2}) z_{ch}- 1 \\
\end{equation}
and $ P_{I}^{I_{3}}$ is the associate Legendre polynomial. The case $ I = I_{3} = 0 $ is particularly simple. The corresponding generating function is 
\begin{equation}
G (z_{0}, z_{ch}) = \int_{0}^{1} \frac{df}{2 \sqrt {2}} \langle e^{\overline {n} z (f) }\rangle \\
\end{equation}
where $ G \equiv G_{00} $ and 
\begin{equation}
z(f) = f z_{0} + (1- f ) z_{ch} - 1.\\
\end{equation}
For simplicity of writing we have left out the $ (s, \vec{b} ) $ dependence in $ \overline {n}(s, \vec{b}) $.

If the classical random source is such that its fluctuations about the mean are Gaussian[27], the generating function becomes 
\begin{equation}
G (z_{0}, z_{ch}) = \int_{0}^{1} \frac{df}{2 \sqrt {2}} e^{\langle \overline{n}\rangle z(f) + \frac{1}{2} d^{2} z^{2}(f)}\\
\end{equation}
where 
\begin{equation}
d^{2} = \langle \overline{n}^{2} (s, \vec{b})\rangle - {\langle \overline {n}(s, \vec{b})\rangle }^{2}.\\
\end {equation}
The form of this generating function is such that the Centauro type behaviour of $ P (n_{0}) $ can be expected only in the limit of a very small dispersion $ d $ in the number of produced pions in b-space. This problem certainly requires further study, in particular, if we go beyond the Gaussian approximation for the fluctuations of the source about the mean.
 
\baselineskip=24pt
\section{Relation to the nonlinear sigma model}

In this section we study the relationship that can be established between the nonlinear sigma model coupled to quarks and our eikonal model with factorization.
\subsection{Quark sources of the pion field}

As is well known, the Lagrangian for QCD with two light up and down quarks has an approximate global 
$SU(2)_{L}\times SU(2)_{R} $ symmetry, which at low temperatures, is spontaneously broken  
to $SU(2)_{V}$ by a nonzero value of the quark condensate $\langle \bar{q}_{L}q_{R}\rangle $,
which is regarded as an order parameter of the system. This order parameter  
can be represented as a four-component vector $\phi \equiv ( \sigma , \vec{\pi })$ buildt from the quark densities.The
chiral symmetry then corresponds to $O($4$)$ rotations in internal space.

The true vacuum of the theory is defined as $ \langle \phi \rangle = ( \langle \sigma \rangle, \vec {0}) $, with $ \langle \sigma \rangle \neq 0 $.
In QCD the spontaneous symmetry breakdown leads to nearly massless Goldstone bosons (the pions) and gives the constituent-quark mass. At low energies and  large distances ( momentum scale smaller than $ 1 GeV $ ) the dynamics of QCD  is described by an effective Lagrangian containing the $\sigma , \vec{\pi} $ fields and constituent quarks. 
Recently, the idea that extended regions of the misaligned chiral field $ \phi $ may be formed in a very high energy hadron-hadron, hadron-nucleus or nucleus-nucleus collisions, has attracted a lot of interest both theoretically and experimentally [17].If a single large DCC domain forms, its decay into pions can lead to anomalous fluctuations in the ratio of neutral to charged pions similar to ones observed in Centauro-type events.

In the DCC dynamics we distinguish three stages: formation, evolution and decay stage. In the conventional approach [16] one starts with a chirally symmetric phase at $ T > T_{c} $ and DCC formation happens as T, due to a rapid expansion or cooling, drops below $ T_{c} $ spontaneously breaking the chiral symmetry.

The evolutionary stage of the DCC is usually described by the classical chiral dynamics based on the $ \sigma $-model, mostly the linear $ \sigma $- model. For the purpose of comparison with our eikonal model we consider the nonlinear $\sigma$-model coupled to quarks at zero temperature [25] which is expected to describe the late stage of the DCC evolution.

The Lagrangian for the nonlinear $\sigma$-model coupled to quarks is 
\begin{equation}
L = \frac{f_{\pi}^{2}}{4} Tr (\partial_{\mu}U^{\dagger} \partial^{\mu} U ) + \overline{q} ( i \gamma \partial) q - g f_{\pi} \overline{q} U q \\
\end{equation} 
where 
\begin{equation}
U = exp ( i \gamma_{5} \frac{\vec{\pi} \cdot \vec{\tau}}{f_{\pi}}) \\
\end{equation}

We parametrize the pion field in the following form 
\begin{equation}
\vec{\pi}(x) = f_{\pi} \vec{n}(x) \theta(x) \\
\end{equation}
where $ \vec{n}(x) $ is an unit vector which determines the isospin orientation of the pion field, obeying $ {\vec{n}}^{2} = 1 $.

The Euler-Lagrange equations of motion for $ \theta $ and $ \vec{n} $ are 
\begin{eqnarray}
\Box \theta - \sin{\theta}\cos{\theta} \partial_{\mu}\vec{n}\cdot \partial^{\mu}\vec{n} & = & -i \frac{m_{Q}}{f_{\pi}^{2}} \vec{n} \cdot (\bar{Q} \vec{\tau} \gamma_{5} Q ) \nonumber \\
\partial_{\mu}( {\sin}^{2}{\theta} \vec{n}\times \partial^{\mu} \vec{n} ) & = & -i \frac{m_{Q}}{f_{\pi}^{2}} \vec{n} \times (\bar{Q} \vec{\tau} \gamma_{5} Q )\sin {\theta} \\
\end{eqnarray}
where Q denotes the constituent quark defined by 
\begin{equation}
Q = e^{i\gamma_{5} \frac {\vec{\pi} \cdot \vec{\tau}}{2 f_{\pi}}} q \\
\end{equation} 
and $ m_{Q} = g f_{\pi} $ is the constituent quark mass. We treat 
\begin{equation}
-i \frac{m_{Q}}{f_{\pi}} \bar{Q} \vec{\tau} \gamma_{5} Q = \vec{j}(x) \\
\end{equation}
as a given classical external source and consider the class of solutions that can be rotated into a uniform one, $ \vec{n}(x) = $ constant, known as the Anselm-class of solutions [13]. The solutions with constant $ \vec{n} $ can be realized if the source points to a certain fixed direction $ \vec{e} $ in the isospace:
\begin{equation}
\vec{j}(x) = j(x) \vec{e} \\
\end{equation}
The source term in the equation of motion for $ \vec{n}$ disappears if we choose 
\begin{equation}
\vec{n}(x) = \vec{e}. \\
\end{equation} 
Then the equation of motion for the pion field reduces to 
\begin{equation}
\Box \theta (x) = j(x) \\
\end{equation}
with 
\begin{equation}
\vec{\pi}(x) = f_{\pi}  \theta(x) \vec{e} \\
\end{equation}

We see that this is exactly the equation of motion for the pion field that we have used in our eikonal model in order to predict the appearance of the Centauro type behaviour in heavy-ion and hadron collisions. This relationship therefore offers the possibility for studying the importance of various quark sources responsible for the DCC formation.

\subsection{Quantum nonlinear  sigma model}

 In quantum chiral field theory  of the nonlinear   $\sigma $-model the role 
of a strong coupling of  $q\pi $-interaction has the quantity $m_{Q}/f_{\pi }$.
However, in phenomenological analyses the strong coupling constant of  $q\pi $-interaction is $g_{Q}$.
It is connected  with $m_{Q}/f_{\pi }$ through the Goldberger-Treiman 
relation  
\begin{equation}
g_{Q} = g_{A}\frac{m_{Q}}{f_{\pi }} \\
\end{equation}
where $g_{A}$  denotes the axial vector current constant whose value, if 
different from unity, should come, according to Lehmann [28], from higher 
orders of perturbation theory. Since the Lagrangian of the nonlinear 
  $\sigma $-model  is nonpolynomial, a suitable renormalization procedure and 
operator normal ordering should be formulated [29]. In this respect, as a first step, we derive the form of the multipion propagator between two quarks. Let 
\begin{equation}
\Phi = :  e^{i\gamma _{5}\frac{\vec{\pi}\vec{\tau}}{f_{\pi}}} - 1 : 
\end{equation}
denote the chiral super field of the pion. The Lagrangian describing the  interaction of 
this super chiral field with quarks is 
\begin{equation}
L  =  \frac{f_{\pi}^{2}}{4}Tr(\partial _{\mu}\Phi \partial ^{\mu}\Phi ) + 
\bar{q}( i \gamma \partial  -  m_{Q} )q  -  \frac{m_{Q}}{f_{\pi}}\bar{q}\Phi q .
\end{equation}
The chiral super propagator  of the field $\Phi $ is defined by 
\begin{equation}
\bigtriangleup _{\Phi }(x) = \langle T(\Phi (x)\Phi (0))\rangle  
\end{equation}
Its form, after number of algebraic manipulations, is 
\begin{equation}
\bigtriangleup_{\Phi }(x) =  \{ 1\otimes 1 + \frac{f_{\pi}^{2}}{3}
(\gamma _{5}\vec{\tau})\otimes (\gamma _{5}\vec{\tau}) \partial _{\triangle}\}
\partial _{\triangle}[\triangle ch(\frac{\triangle }{f _{\pi}^{2}})] + 1
\end{equation} 
where 
\begin{equation}
\langle T(\pi _{i}(x)\pi _{j}(0))\rangle  =  \delta _{ij} \triangle (x).
\end{equation}
and \\
\[ \triangle (x) =  \frac{1}{4\pi ^{2}}\frac{1}{x^{2} - i\epsilon }\]. \\
The multipion exchange potential between two quarks is related to the 
Fourier transform of $\triangle _{\Phi }(x)$ in the following way:
\begin{equation}
\bar{u}(p_{1}')\bar{u}(p_{2}') \tilde{\triangle }_{\Phi }(q) u(p_{1})u(p_{2}) = 
\omega '_{1}\omega '_{2} \tilde{V}_{\Phi }(q)\omega _{1}\omega _{2}
\end{equation}
where \\
\[ u(p) = \sqrt{\frac{\epsilon _{p} + m}{\epsilon _{p}}}
{\omega \choose \omega '}\] \\
and \\
\[ \omega ' = \frac{\vec{\sigma }\vec{p}}{\epsilon _{p} + m}\omega \] \\
with $\omega ^{*}\omega  = 1$.

We should also mentioned  that the soft chiral pion bremsstralung [30-33], 
in which every incoming and outgoing quark line is replaced by 
\begin{equation}
q \longrightarrow  exp (i\gamma _{5}\frac{\vec{\pi }\vec{\tau }}{2f _{\pi }}) q
\end{equation}
also leads  to the distribution of neutral pion of the form  $1/\sqrt{n_{0}n}$ which
is typical for coherent pion production without invoking the notion of DCC formation.
We  conclude therefore that  $n P(n_{0}\mid n) \sim \sqrt{n/n_{0}}$ is not always a definite 
signature of DCC formation.

\section{Conclusion}
The result of our analysis are the following:
\begin{itemize}
     \item Within the framework of an unitary eikonal model with 
               factorization and global conservation of isospin the Centauro-type 
               behaviour can only be expected for pions which are produced singly 
               ( see Fig.1 and 2 ).
     \item  The Centauro-type effect depends on isospin of the initial-leading-
                particle system (see Fig.1).
      \item  The coherent production of  $\rho $-type clusters supresses the 
                 Centauro-type behaviour (see Fig. 2).
      \item  A classical random pion source, which has Gaussian fluctuations about 
                 the mean, also supresses the Centauro-type behaviour.
       \item  The  leading-particle effect and the factorization property of 
                  the scattering amplitude in the impact parameter space may 
                  be related to the isospin-uniform solutions of the nonlinear 
                  $\sigma $-model coupled to quarks.
       \item   The multipion exchange potential between two quarks is 
                   derived which is different from the one-gluon exchange potential. 
        \item   The soft chiral pion bremsstralung  also leads to anomalously 
                    large Centauro-type ratio of neutral to charged pions.
\end{itemize}

\baselineskip=24pt
{\large \bf Acknowledgment }
This work was supported by the Ministry of Science of
Croatia under Contract No. 0098104.

\newpage

{\bf Figure captions :}

Fig. 1. The distribution of neutral pions  as a function of
the isospin of the incoming leading-particle system in comparison with the binomial distribution.
 
Fig. 2. The curves represent $ P (n_{0}) $ for different combinations of
$( \overline{n}_{ \pi},  \overline{n}_{ \rho}),$
the average number of singly produced pions and the average
number of $ \rho$-type clusters, respectively.

\end{document}